\documentclass[openacc]{rstransa}

\newcommand{\lenscat}[1]{%
  {\texttt{lenscat}#1}}
\usepackage{tabularx}
\usepackage{moresize}

\titlehead{Research}

\begin{document}

\title{\lenscat: a Public and Community-Contributed Catalog of Known Strong Gravitational Lenses}

\author{
L. Vujeva, R. K.~L.~Lo, J. M. Ezquiaga, \\
J. C.~L.~Chan}


\address{Center of Gravity, Niels Bohr Institute, Blegdamsvej 17, 2100 Copenhagen, Denmark}


\subject{physics, astrophysics, galaxies, cosmology, observational astronomy}

\keywords{gravitational lensing, gravitational waves, gamma ray bursts, kilonovae, supernovae, radio transients}

\corres{Luka Vujeva\\
\email{luka.vujeva@nbi.ku.dk}}

\begin{abstract}
We present \lenscat, a public and community-contributed catalog of strong gravitational lenses found by electromagnetic surveys. 
The main objective of \lenscat~is to compile a simple, easy-to-access catalog that can be used in a variety of lensing studies, such as facilitating the search for the host galaxy of a candidate strongly lensed transient event. We also provide a python package to interact with tools commonly used by the community.
This allows end users both with and without lensing expertise to obtain a list of known strong lenses within a given search area, and to also rank them by their respective searched probabilities. 
Here, we exemplify this by crossmatching the gravitational wave joint sky localization region of an interesting pair of events GW170104-GW170814. Other examples including cross-matching short gamma-ray bursts are given.  
Thanks to the open and simple infrastructure of \lenscat, members of the lensing community can directly add newly found lenses from their own studies to help create a long-lasting catalog that is as exhaustive and accessible as possible. 
\end{abstract}


\maketitle


\section{Introduction}

Gravitational lensing has a rich history in the electromagnetic (EM) spectrum, and has provided us with great insights about the structure of dark matter halos in galaxies and galaxy clusters \cite{bartelmann2010review}, as well as the large scale structure of the Universe. 
These observations correspond mostly to static or variable sources, such as galaxies or quasars, whose emitted light is lensed by cosmic structures such as galaxies or galaxy clusters. 
By now we accumulate thousands of such observations. 
However, we have just started to detect strongly lensed transients (namely 8 supernovae \cite{nikkisnlens, sn1, sn2, sn3, sn4, sn5, sn6, sn7, sn8}), and we have not yet observed any lensed multi-messenger event. 
Finding lensed transients is the next frontier in gravitational lensing and a strongly lensed multi-messenger event would be an invaluable probe of general relativity as well as cosmology \cite{bartelmann2010review,blandfordnarayanreview,oguri2019review,liao2022review,sahareview2024}. Given the rarity of strong lensing \cite{robertson2020, ligo03halflensing, ld2020} in combination with current detector sensitivities (e.g. for gravitational wave (GW) detectors, see \cite{advligo, advligo2, kagrarev}), finding a strongly lensed multi-messenger transient will not be easy.

The successful identification of the host galaxy of a strongly lensed transient is heavily dependent on having a complete, publicly available catalog of all known strong gravitational lenses from electromagnetic surveys in order to ensure that all possible lens candidates are being considered. 
Given the limited amounts of observing time, \lenscat\ can be used to prioritize deeper observations of likely candidates within the large $10 - 10^4$ square degree sky localization regions that are typical of binary black hole mergers \cite{ligoobsscen}. This will also aid in faster host identification and follow-ups for lensed transients such as binary neutron stars, allowing the community to take detailed spectroscopy of the event as quickly after the merger as possible.

Although there are some databases that contain many lenses, such as Master\footnote{\url{https://test.masterlens.org/index.php}}\cite{moustakas2012}, most lens catalogs are either private within collaborations, or are not being actively maintained. The \lenscat\ catalog, a completely public catalog to be presented in this paper, was created to store a simple list of known strong lensing galaxies and clusters, and also to provide an easy-to-use platform that future surveys with wide field telescopes such as the Vera Rubin Observatory (VRO) \cite{zeljkolsst} and Euclid \cite{euclid1} can use to store their newly found lenses for the rest of the community to use. 
Moreover, given the explosion of the number of transients that we will observe in the near future (e.g. $\sim10^7$ with VRO during its ten year survey \cite{lsstbook} and $\sim10^3$ per year with current GW detectors at their design sensitivity \cite{ligoobsscen}), validating candidates with known lens catalogs will be even more important.

The paper is organized as follows: Section~\ref{sec:catalog} will describe the catalog and its features, Section~\ref{sec:crossmatch} will show an example of the catalog in use, and Section~\ref{sec:future} will discuss the future science goals that \lenscat\ will facilitate.

\section{The \lenscat\ catalog}
\label{sec:catalog}

\begin{figure}[!h]
\centering\includegraphics[width=1.\linewidth]{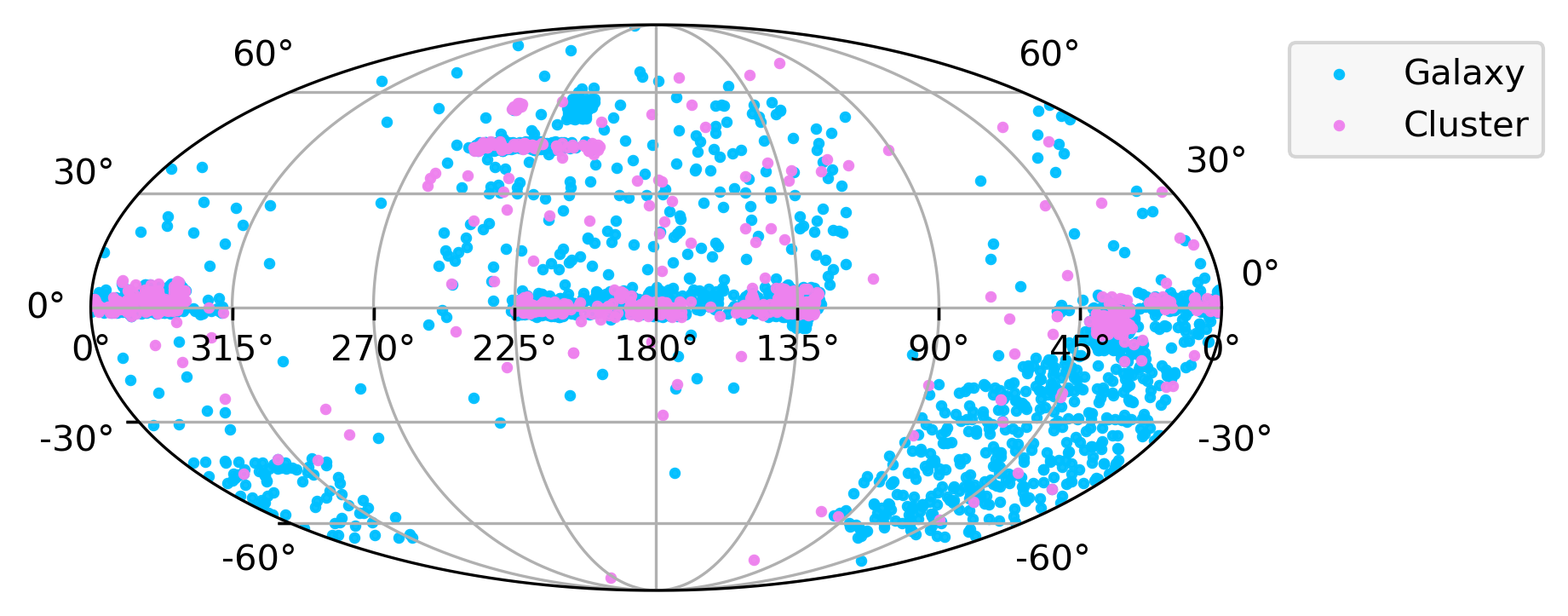}
\caption{Spatial distribution of the lensing galaxies (blue) and galaxy clusters (pink) in \lenscat. Currently, there are 4587 known lenses included in our catalog. 
The lack of found lenses corresponds mostly to the galactic plane owing to the lack of suitable quality of survey data.}
\label{catalog}
\end{figure}

The \lenscat\ catalog is a compilation of existing publicly available strong lens catalogs \cite{3,4,5,6,7,8,9,10,11,12,13,14,17,15,16}. Only the essential quantities associated with each lens are kept in our catalog, and they are listed in Table \ref{tab:candidates}. A notable exception is that the source redshift associated with each lensing system is excluded because only a small subset of the entries in the catalog have measurements of the source redshift. In addition, systems such as cluster lenses can have many source redshifts associated with a single lens.
For further information, the reference associated with each lens is provided so that the user interested in more detailed analyses can easily access all the information.\footnote{A complete list of the studies from which we drew lenses from can be found on the project's GitHub page \url{https://github.com/lenscat/lenscat?tab=readme-ov-file}. The list is continually updated as new lenses are added to the catalog.}

\begin{table}[!h]
\caption{List of the parameters describing each lens in the catalog.}
\label{tab:candidates}
\begin{tabular}{|p{1.5in}|p{3.in}|}
\hline
Column Name & Description \\
\hline
\hline
\texttt{name} & Names of galaxies/galaxy clusters \\
\hline
\texttt{RA [deg]} & Right ascension in degrees \\
\hline
\texttt{DEC [deg]} & Declination in degrees \\
\hline
\texttt{zlens} & Lens redshift (if known) \\
\hline
\texttt{type} & Type of lens (i.e. galaxy or galaxy cluster) \\
\hline
\texttt{grading} & Grading whether it is a "confident" lens or a "probable" lens (see following section for explanation of the grading scheme) \\
\hline
\texttt{ref} & Reference to the corresponding catalog or study   \\\hline
\end{tabular}
\vspace*{-4pt}
\end{table}

An important variable of any lens catalog is the grading assigned to each lens. Those, however, are typically very different across different surveys. 
To minimize this variance, the lens grading that is being employed in this catalog is a simple convention that is determined by the nomenclature of the contributor's study. 
For example, where some members of the community will only deem a lens "confident" if the multiple images associated with the system have been spectroscopically confirmed, machine learning based studies will give this distinction to the lenses that are given the top grade in their study. Because of the lack of unified definitions, the grading given in \lenscat\ is simply a quick reference for the user, and is not to be taken as the absolute truth about the system. The in-depth lens grading systems created by the individual contributors can be seen in the references linked with every lens in the catalog, which should provide end users with added information should they require it.
The \texttt{lenscat} catalog defines a "confident" lens as either having spectroscopically confirmed multiple images, or to be within the highest grading of the study presenting the lenses. Should the community disagree with the classification of a lens, they are encouraged to submit a pull request to the GitHub repository to change the classification of the lens.

There are three main ways to use \lenscat. The first is through the python package\footnote{\url{https://pypi.org/project/lenscat/}}, which contains the catalog, as well as a suite of tools designed to facilitate common use cases for the catalog (which will be further discussed in Section~\ref{sec:crossmatch}). The three main features are basic searching with \texttt{.search()}, crossmatching with a skymap with \texttt{.crossmatch()}, and visualizing with \texttt{.plot()}. Note that these functions will return a Catalog object, and hence they can be composed together (e.g., \texttt{.crossmatch().search())}. For quick searches, one can use the web interface\footnote{\url{https://lenscat.streamlit.app/}}, which has many of the functionalities present in the python package, but is presented in a beginner friendly manner, without having to write a single line of code. Finally, the catalog is also available as a plain csv file for those that wish to code in their preferred programming language.

\section{Crossmatching}
\label{sec:crossmatch}

One of the core features of \lenscat\ is  crossmatching, which should facilitate validating and finding the host galaxy, galaxy group, or galaxy cluster of lensed transients. 
Crossmatching will also be essential for the search for multi-messenger counterparts of candidate lensed transients. 
This will be especially relevant for lensed GWs and their potential EM signals. Using the available information about each lens within a given credible region of the sky can allow for additional consistency checks between the image properties of the EM photometry/spectroscopy of the multiple images, and the GW observables such as time delays and relative magnification factors~\cite{otto2019,janquarto3, wempe2022, magare2023}. This will allow the EM followups to prioritize observing the lens systems that best match these consistency checks.

There are a multitude of options within the python package that allow users to customize their searches. In this section, we demonstrate an example of taking the joint skymap of two gravitational-wave events that are hypothetically two strongly lensed images of the same source, and finding known strong lenses that fall within a given credible region. This feature is implemented as \texttt{.crossmatch()}. This function is simply a wrapper to the \texttt{crossmatch()} function in \texttt{ligo.skymap}\footnote{\url{https://git.ligo.org/lscsoft/ligo.skymap}} which performs the cross-matching of a GW skymap with a given list of coordinates from the lens catalog. 

\begin{figure}[!h]
\centering\includegraphics[width=0.63\linewidth]{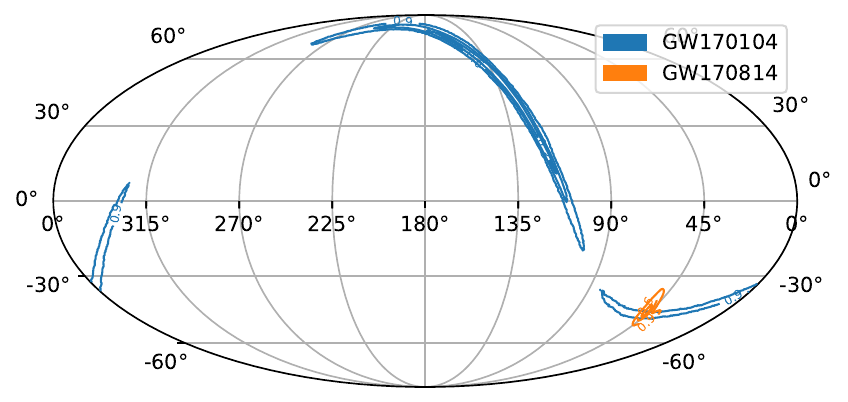}
\centering\includegraphics[width=0.33\linewidth]{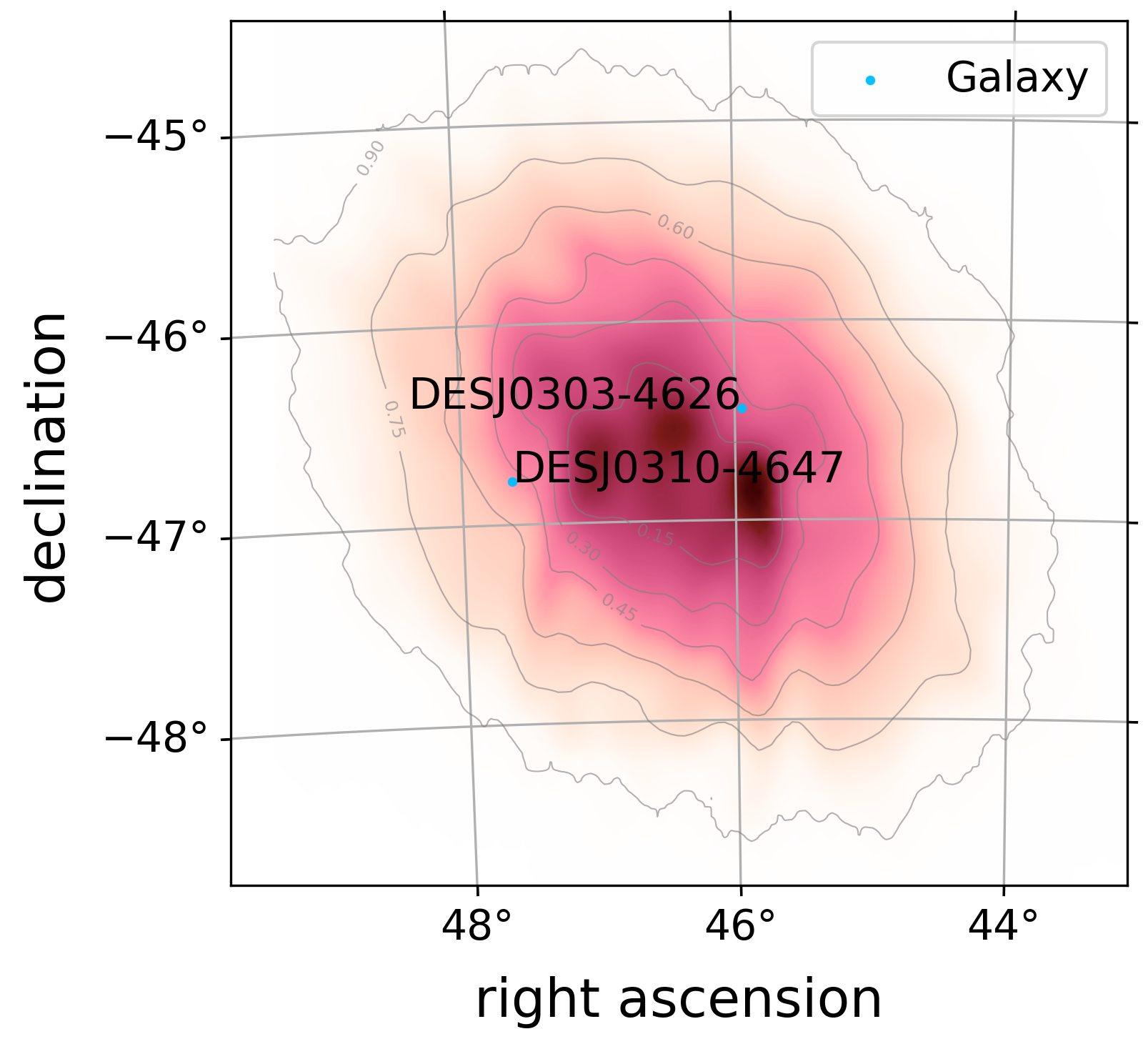}
\caption{Skymaps of GW170104 and GW170814 (left), and their joint skymap crossmatched with \texttt{lenscat} (right).}
\label{fig:crossmatch}
\end{figure}

As a simple case of study we consider now an interesting pair of gravitational-wave events GW170104-GW170814,
which have overlapping sky localizations \cite{ricooverlap} as shown in the left hand side of Fig. \ref{fig:crossmatch}. Although these two events have been deemed unlikely to be a real lensed pair of images \cite{ligo03halflensing, otto2019, ld2020, liu2021, ricooverlap}, we will simply use them to show the capabilities of \texttt{lenscat}.
For example, to cross-match the joint GW skymap of GW170814 \cite{gw170814} and GW170104 \cite{gw170104} with only the confident galaxy lenses in \lenscat, simply run \texttt{lenscat.catalog.search(lens\_type= \\ "galaxy").crossmatch("joint\_skymap.fits.gz")}. 
The result is plotted in the right hand side of Fig. \ref{fig:crossmatch}.

This returns a catalog of the lenses found within a given credible region. 
The (truncated) output of such search is displayed below:

\

\ssmall
\begin{verbatim}
<CrossmatchResult length=4587>
       name             RA          DEC      zlens     type   grading  searched probability   searched area   
                       deg          deg                                                            deg2       
      str20          float64      float64    str15     str7     str9         float64             float64      
------------------ ------------ ----------- -------- ------- --------- -------------------- ------------------
     DESJ0303-4626      45.9507   -46.44066     1.37  galaxy  probable   0.2704101345085984 1.4195865170954403
     DESJ0310-4647     47.63526   -46.78398     0.71  galaxy  probable    0.504649447144695  3.258819856796524
\end{verbatim}
\normalsize

In this case one can see that there are only 2 galaxy lenses within the 95\% credible region. Note that due to catalogue incompleteness, the two lenses shown within the skymap are not guaranteed to be the only lenses in this region, and cannot be guaranteed to be the host galaxies associated with the corresponding lensed event.

It is important to emphasize that \lenscat\ is not only a tool for lensed gravitational waves, but is compatible with almost any transient skymaps (i.e. Gamma Ray Burst (GRB), Fast Radio Burst (FRB), etc.). For example, one could crossmatch the skymap of "GRB 240229A" simply by downloading the skymap in a FITS file format\footnote{The public skymap can be found at \url{https://heasarc.gsfc.nasa.gov/FTP/fermi/data/gbm/triggers/2024/bn240229588/quicklook/glg_healpix_all_bn240229588.fit}} and running \texttt{lenscat.catalog.crossmatch("grbskymap.fits")}, where \texttt{grbskymap.fits} is the name of the skymap associated with "GRB 240229A". 
In this case one would find 5 lenses within the 70\% confidence region.

\section{Future Prospects}
\label{sec:future}

The primary purpose of \lenscat \ is to facilitate follow-up searches for lensed transients. As mentioned in earlier sections, this can be applied to a variety of transients, such as gravitational waves, gamma ray bursts, and fast radio bursts. With a complete enough lens catalog, follow up observations can prioritize covering the lensed lines of sight in their search region, which can speed up the process of finding a lensed EM counterpart. This will be further facilitated by how simple \lenscat \ is to use, which will allow for both experts and newcomers to the field alike to obtain the information they need from the catalog as quickly as possible, leading to faster follow up times. 

However, the success of identifying the host galaxies of strongly lensed gravitational waves will be strongly dependent on the completeness of the EM catalog. Hence, \lenscat\ is a community-contributed and maintained catalog, meaning that any member of the lensing community can submit new lenses, or replace information about current lenses with updated information. This procedure is done entirely through GitHub pull requests, which will be approved by moderators. Making this process as streamlined as possible is key for the success of such a catalog, because the easier it is to submit new lenses to the catalog, the more likely members of the community are to contribute. Community members can also contribute to the project by moderating the submission of new lenses or changes to information associated with currently known lenses.  

Currently, cluster scale lenses are defined as falling within a specified credible region of a skymap based on whether their centroid falls within this region. However, the strong lensing regions of clusters can be extended, meaning that part of the region of interest might lie outside of the credible region. A future version of the code could increase the region in which a cluster is considered to be part of of a credible region of a skymap based on its mass or Einstein radius in order to ensure that clusters just outside of a credible region are still properly considered as being part of the known lenses in the credible region.

Community-contributed lens catalogs
will be especially important with the large number of new lenses that will be discovered with upcoming wide field telescopes such as the Vera Rubin Observatory or Euclid. The lenses discovered by these telescopes will be key in building a complete catalog because of their combination of depth, wide search area, and high cadence (which will allow for the discovery of more lensed transients). Having all of the newly discovered lenses in one open access catalog will be important for the wider astronomy and astrophysics community to gain as much from the data as possible. Having all known lenses in one catalog will also facilitate studies of both lens populations and the populations of lensed images, which will provide insight into the dark matter halo distribution in the Universe, as well as the properties of the dark matter halos.

\ack{The authors would like to thank Anupreeta More and Masamune Oguri for their helpful comments and insights. 
The Center of Gravity is a Center of Excellence funded by the Danish National Research Foundation under grant No. 184.
This project was supported by the research grant no. VIL37766 and no. VIL53101 from Villum Fonden, and the DNRF Chair program grant no. DNRF162 by the Danish National Research Foundation.
This project has received funding from the European Union's Horizon 2020 research and innovation programme under the Marie Sklodowska-Curie grant agreement No 101131233.  
JME is also supported by the Marie Sklodowska-Curie grant agreement No. 847523 INTERACTIONS. 
The Tycho supercomputer hosted at the SCIENCE HPC center at the University of Copenhagen was used for supporting this work. This material is based upon
work supported by NSF's LIGO Laboratory which is a major facility fully funded by the
National Science Foundation. LV would also like to thank Samsung DeX for facilitating the writing of this manuscript.
}


\bibliography{references}
\end{document}